\PassOptionsToPackage{table}{xcolor}
\documentclass[conference]{IEEEtran}

\usepackage{algorithm}
\usepackage[noend]{algpseudocode}
\usepackage{amsmath}
\usepackage{amssymb}
\usepackage{graphicx}
\usepackage{url}
\usepackage{listings}
\usepackage{subcaption}
\usepackage{tikz}
\usepackage{cite}
\usepackage[inline]{enumitem}
\usepackage{hyperref}
\usepackage{tabularx}
\usepackage{multirow}
\usepackage{multicol}
\usepackage{textgreek}
\usepackage{tablefootnote}
\usepackage{threeparttable}
\usepackage{fancyhdr}
\usepackage[table]{xcolor}
\hypersetup{
  colorlinks,
  allcolors=blue,
  citecolor=blue,
  linkcolor=blue,
  urlcolor=blue
}
\usepackage{xcolor}
\usepackage{circledsteps}

\newcommand\myCircled[2][]{%
  \ifmmode
    \Circled[fill color=black, inner color=white, #1, inner xsep=0.5ex, inner ysep=0.5ex]{\scriptsize\mathsf{#2}}%
  \else
    \Circled[fill color=black, inner color=white, #1, inner xsep=0.5ex, inner ysep=0.5ex]{\scriptsize\sffamily#2}%
  \fi
}

\begin{document}

\title{TLGLock: A New Approach in Logic Locking Using Key-Driven Charge Recycling in Threshold Logic Gates}

\author{\IEEEauthorblockN{Abdullah Sahruri}
\IEEEauthorblockA{\textit{School of Computing and Informatics}\\
\textit{University of Louisiana at Lafayette}\\
Lafayette, Louisiana, USA\\
abdullah.sahruri1@louisiana.edu}
\and
\IEEEauthorblockN{Martin Margala}
\IEEEauthorblockA{\textit{School of Computing and Informatics}\\
\textit{University of Louisiana at Lafayette}\\
Lafayette, Louisiana, USA\\
martin.margala@louisiana.edu}
}

\maketitle
\renewcommand{\headrulewidth}{0.0pt}
\thispagestyle{fancy}
\lhead{}
\rhead{}
\chead{This is the author's version of the work. The definitive Version of Record will appear in the 2025 IFIP/IEEE International Conference on Very Large-Scale Integration (VLSI-SoC).}
\cfoot{}
\begin{abstract}
Logic locking remains one of the most promising defenses against hardware piracy, yet current approaches often face challenges in scalability and design overhead. In this paper, we present \textbf{TLGLock}, a new design paradigm that leverages the structural expressiveness of Threshold Logic Gates (TLGs) and the energy efficiency of charge recycling to enforce key-dependent functionality at the gate level. By embedding the key into the gate's weighted logic and utilizing dynamic charge sharing, TLGLock provides a stateless and compact alternative to conventional locking techniques. We implement a complete synthesis-to-locking flow and evaluate it using ISCAS, ITC, and MCNC benchmarks. Results show that TLGLock achieves up to \textbf{30\% area}, \textbf{50\% delay}, and \textbf{20\% power} savings compared to latch-based locking schemes. In comparison with XOR and SFLL-HD methods, TLGLock offers up to \textbf{3× higher SAT attack resistance} with significantly lower overhead. Furthermore, randomized key-weight experiments demonstrate that TLGLock can reach up to \textbf{100\% output corruption} under incorrect keys, enabling tunable security at minimal cost. These results position TLGLock as a scalable and resilient solution for secure hardware design.
\end{abstract}


\section{Introduction}

Logic locking is an essential technique for protecting integrated circuits (ICs) against reverse engineering and unauthorized use. The globalization of the IC supply chain, driven by cost reduction and rapid time-to-market, has amplified risks such as intellectual property (IP) piracy, overproduction, and unauthorized activation, especially in the presence of third-party vendors \cite{tehranipoor2010survey}. Traditional logic locking schemes like XOR/XNOR-based insertion offer baseline security but are vulnerable to SAT-based attacks that can efficiently recover secret keys \cite{yasin2016sarlock}.

To address these vulnerabilities, resilient schemes like SARLock and Anti-SAT were proposed \cite{yasin2016sarlock, xie2018anti}. However, these methods significantly increase area, delay, and power due to additional logic and key-validation circuitry. This trade-off between security and circuit performance poses a challenge, particularly for area- and power-constrained systems. Furthermore, techniques such as point function obfuscation, scan chain locking, and routing-based methods face challenges in output corruptibility, scalability, and design complexity \cite{yasin2017provably}.

\begin{figure}[t!]
\centering
\includegraphics[width=1\linewidth]{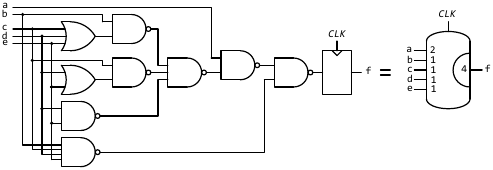}
\caption{A single Threshold Logic Gate (TLG) replacing a complex multi-level CMOS logic network.}
\label{fig:tlgdepth}
\vspace{-1em}
\end{figure}

Threshold Logic Gates (TLGs) offer a promising alternative for secure and efficient digital logic synthesis. TLGs, as seen in Fig.~\ref{fig:tlgdepth}, replace multi-level CMOS logic with a single gate by computing a weighted sum of binary inputs and comparing it to a threshold:

\begin{equation}\label{eq:y1}
    \sum_{j=1}^{m} X_j \times W_j > T \quad \Rightarrow \quad Z=1
\end{equation}
\begin{equation}\label{eq:y0}
    \sum_{j=1}^{m} X_j \times W_j \leq T \quad \Rightarrow \quad Z=0
\end{equation}

where $X_j \in \{0,1\}$ are binary inputs, $W_j$ are integer weights, and $T$ is the threshold. This compact representation improves logic density and can be exploited for both performance and obfuscation.

More than 50 TLG circuit designs have been documented \cite{beiu2003threshold}, and they have been used in arithmetic logic units (ALUs) \cite{medina2019reconfigurable}, processing-in-memory systems \cite{ angizi2019aligns, angizi2020exploring}, and low-power architectures \cite{he2016energy, papandroulidakis2019practical,sahruri2024hictl}. TLGs can be implemented using various CMOS techniques, including traditional, conductance-based, and capacitive structures \cite{ozdemir1996capacitive,lopez2004balanced,leblebici1996compact}. Among these, Latch-type Threshold Logic (LCTL) \cite{avedillo1995low} and Charge Recycling Threshold Logic (CRTL) \cite{celinski2001low} offer contrasting trade-offs: \textit{LCTL} supports high fan-in designs with low static power but incurs higher area, while \textit{CRTL} achieves superior energy efficiency and speed due to its charge recycling operation.

These features make TLGs naturally suited for logic locking. By embedding key inputs into the weighted structure of TLGs, one can obscure functionality without adding external logic gates. However, the use of threshold-based locking has not been thoroughly explored in terms of its synthesis flow, key embedding mechanism, or trade-offs in power, delay, and area.

\begin{figure}
    \centering
    \includegraphics[width=1\linewidth]{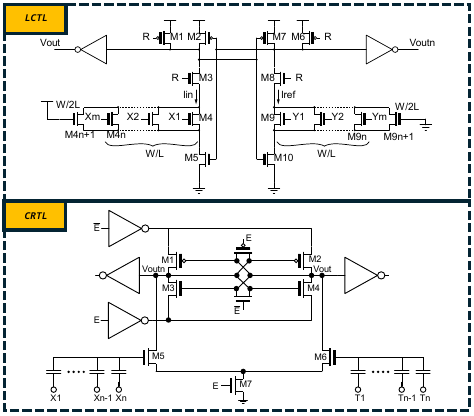}
    \caption{TLG architectures: Latch-type Low Power Threshold Logic (\textit{LCTL}) and Charge Recycling Threshold Logic (\textit{CRTL}).}
    \label{fig:ctl}
\end{figure}

In this work, we introduce a threshold-locked design methodology that leverages both \textit{LCTL} and \textit{CRTL} gates, as illustrated in Fig.~\ref{fig:ctl}, and implemented using Cadence’s 45nm GPDK045 technology. Our approach integrates secret keys directly into the gates by mapping them as input variables with tunable weights, enabling a native, compact, and efficient locking mechanism. To support this methodology, we develop a complete synthesis flow and evaluate its effectiveness using standard benchmark suites, including ISCAS’85, ISCAS’89, and ITC’99.

Our results show that \textit{CRTL}-based locking achieves up to \textbf{30\% area}, \textbf{50\% delay}, and \textbf{20\% power} reduction over its \textit{LCTL} counterpart. Moreover, SAT solvers fail to break large designs like \texttt{c7552}, \texttt{i10}, and \texttt{b17} within a one-hour timeout, demonstrating strong attack resilience. The proposed technique outperforms conventional locking methods in both efficiency and security.

The rest of the paper is organized as follows: Section~\ref{sec:overheads} discusses limitations in traditional logic locking. Section~\ref{sec:flow} introduces our TLG-based locking flow. Section~\ref{sec:results} presents experimental analysis on area, power, delay, and SAT resilience. Section~\ref{sec:conclusion} concludes the work.

\begin{figure*}[ht!]
    \centering
    \includegraphics[width=0.9\linewidth]{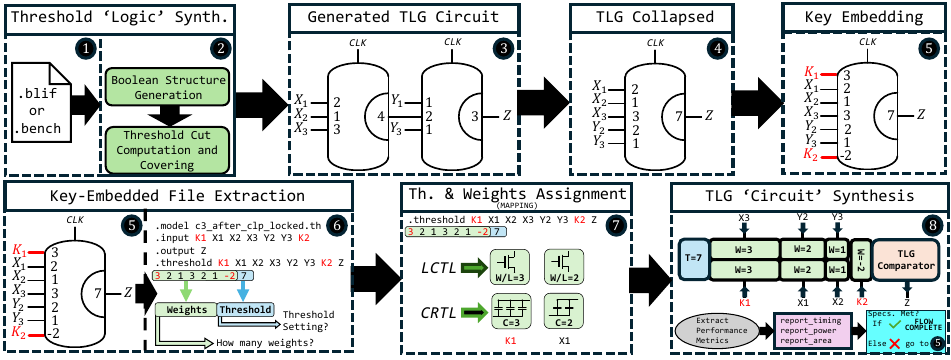}
    \caption{Proposed flow: \myCircled{1} Threshold logic synthesis from a BLIF or BENCH file involving Boolean structure analysis and threshold cut computation. \myCircled{2} TLGs with respective weights and thresholds post-synthesis. \myCircled{4} Merging two TLGs via linear combination \cite{lee2016analytic}. \myCircled{5} Embedding key inputs with weights proportional to the input weight sum. \myCircled{6} Extraction of weights and threshold values to be used in \myCircled{7} for assigning them into their corresponding inputs.\myCircled{8} TLG circuit synthesis and PnR (Place and Route) of the TLG circuits for performance extraction.}
    \label{fig:flow}
    \vspace{-1em}
\end{figure*}

\section{Overhead in Logic Locking Techniques}
\label{sec:overheads}


Early XOR/XNOR-based locking offered low overhead but was quickly shown to be vulnerable to SAT-based attacks, which efficiently recover secret keys using oracle access to the unlocked design \cite{roy2010epic, subramanyan2015evaluating}. To mitigate these vulnerabilities, SAT-resilient methods such as SARLock, Anti-SAT, and DisORC introduced structural redundancy and point function obfuscation, but they significantly increased area and power \cite{yasin2016sarlock, Limaye2021thwarting}.

Compound locking schemes emerged to combine the benefits of multiple methods. For instance, integrating XOR/XNOR logic with point function obfuscation enhances resilience but further inflates design overhead due to the added programmable gates and logic layers \cite{kamali2022advances}.

Cyclic-based approaches further complicate SAT attacks by embedding feedback loops, increasing attack complexity exponentially with the number of cycles. However, they often incur high area costs and design rule challenges \cite{roshanisefat2018srclock, rezaei2019cycsat}.

Emerging techniques involving LUTs and FSM-based obfuscation offer implicit key embedding and obfuscated functionality, but these often require additional logic states or large memory-like blocks, contributing to significant power and area penalties. In testable designs, scan chains also pose a threat by exposing internal signals and key values.

In summary, while traditional logic locking methods offer various degrees of security, they come with inherent trade-offs—either high performance overhead or limited attack resilience. These limitations motivate the need for alternatives that can offer strong protection with compact, efficient hardware. In the next section, we present a new TLG-based locking methodology that embeds key functionality directly into the threshold logic synthesis process—achieving robustness and performance simultaneously.

\section{Proposed Flow}
\label{sec:flow}
The proposed flow, depicted in Fig.~\ref{fig:flow}, outlines the process of embedding logic locking within threshold logic gates synthesized from And-Inverter Graphs (AIGs). This section outlines how key inputs are integrated into the threshold logic structure to ensure secure circuit locking. The proposed flow for the synthesis and design of threshold logic circuits begins with \myCircled{1}, where the synthesis starts by analyzing the given Boolean structure from either a BLIF or BENCH file format. This stage involves a comprehensive examination of the logic structure to compute threshold cuts that are capable of representing the Boolean functions using TLGs \cite{neutzling2018effective}. In \myCircled{2}, after synthesis, TLGs are produced, each characterized by their specific weights and threshold values based on the input requirements. Moving to \myCircled{4}, the process supports merging two TLGs through linear combination, leveraging analytic techniques as outlined in prior works \cite{lee2016analytic}. This merging strategy helps optimize the circuit structure by combining gates efficiently, thereby simplifying the overall design.

In \myCircled{5}, key inputs are embedded into the design, with their assigned weights set proportionally to the sum of the input weights, ensuring the circuit's functionality and security through logic locking mechanisms. This step adds an extra layer of protection by integrating keys in a way that leverages the threshold characteristics of the gates. Stage \myCircled{6} focuses on the extraction of weights and threshold values from the previously synthesized and merged TLGs, preparing these critical parameters for assignment in \myCircled{7}. Here, the extracted weights and thresholds are accurately mapped to their respective inputs, ensuring the correct behavior of the TLG-based design.

Finally, in \myCircled{8}, the flow concludes with the synthesis of the TLG circuit. This step is essential for physical design verification, where the circuit's performance is extracted and assessed. The process iterates until the performance meets the predefined specifications, ensuring that the TLG circuit design aligns with desired delay, area, and power metrics.
\begin{algorithm}[htbp]
\caption{Threshold Logic Synthesis and Key Preparation}
\label{alg:TLG_synthesis}
\begin{algorithmic}[1]
\Require Original circuit $C_{\text{orig}}$, \#Keys $N_k$, Insertion \% $P$
\Ensure Synthesized circuit $C_{\text{TLG}}$, Key vector $K$
\State $C_{\text{TLG}} \gets$ \textsc{SynthesizeTLG}$(C_{\text{orig}})$
\State Select $\mathcal{G}_{\text{lock}} \subseteq C_{\text{TLG}}$ with $|\mathcal{G}_{\text{lock}}| = \lceil |C_{\text{TLG}}| \cdot \frac{P}{100} \rceil$
\State $K \gets \{k_1, \ldots, k_{N_k}\}$ \Comment{Random key vector}
\State \Return $C_{\text{TLG}}, K$
\end{algorithmic}
\end{algorithm}

\begin{algorithm}[htbp]
\caption{Key Integration and Logic Locking}
\label{alg:Key_integration}
\begin{algorithmic}[1]
\Require TLG circuit $C_{\text{TLG}}$, Key vector $K$, Thresholds $T_j$
\Ensure Locked circuit $C_{\text{locked}}$
\ForAll{$g_j \in \mathcal{G}_{\text{lock}}$}
    \State Integrate key bits $\{k_{j1}, \dots, k_{jm}\}$ with weights $\{v_{j1}, \dots, v_{jm}\}$
    \State Update sum: $S_j = \sum w_i x_i + \sum v_{jl} k_{jl}$
    \State Set $g_j.\text{output} \gets \mathbb{1}[S_j \geq T_j]$
\EndFor
\State \Return $C_{\text{locked}} \gets$ \textsc{Format}$(C_{\text{TLG}})$
\end{algorithmic}
\end{algorithm}
To further illustrate the steps in this flow, the detailed TLG synthesis and locking processes are formalized in Algorithm~\ref{alg:TLG_synthesis} and Algorithm~\ref{alg:Key_integration}. Algorithm~\ref{alg:TLG_synthesis} describes the initial synthesis of the original circuit netlist $C_{orig}$ into a TLG-based circuit $C_{TLG}$. It calculates the properties of each gate, including weights ($w_i$), fan-in ($f_{in}$), and fan-out ($f_{out}$), and determines the number of gates to be modified for key integration based on a specified percentage $P$. A key vector $K$ is generated to facilitate the embedding of security elements into the circuit.

Algorithm~\ref{alg:Key_integration} focuses on embedding the key inputs within the synthesized TLG circuit $C_{TLG}$. For each selected gate $g_j$, key inputs $\{k_{j1}, k_{j2}, \ldots\}$ are integrated with corresponding weights, and the combined weighted sum is evaluated against a threshold value $T_j$. The output of the gate is set based on whether the total weighted sum meets or exceeds $T_j$. This process ensures that the resulting locked circuit $C_{locked}$ relies on the correct key for proper functionality, providing an effective security mechanism. Finally, the circuit is formatted and validated to produce the locked netlist $C_{locked}$.

The integration process described in Algorithm~\ref{alg:Key_integration} is mathematically formalized by extending the standard weighted sum equation of TLGs to incorporate key inputs. The traditional operation of a TLG is defined as:

\begin{equation}
\text{Output} = 
\begin{cases} 
1, & \text{if } \sum_{i=1}^{n} w_i \cdot x_i \geq T \\
0, & \text{otherwise,}
\end{cases}
\end{equation}

where \( x_i \) are the input signals, \( w_i \) are their associated weights, and \( T \) is the threshold value. 

To integrate logic locking, we introduce key inputs \( k_j \) with corresponding weights \( v_j \). The modified equation becomes:

\begin{equation}
\text{Output} = 
\begin{cases} 
1, & \text{if } \sum_{i=1}^{n} w_i \cdot x_i + \sum_{j=1}^{m} v_j \cdot k_j \geq T \\
0, & \text{otherwise.}
\end{cases}
\end{equation}

This modification ensures that the circuit produces the correct output only when the appropriate key vector \( \mathbf{K} = [k_1, k_2, \ldots, k_m] \) is applied. An incorrect key vector \( \mathbf{K}' \) introduces a deviation in the weighted sum:

\begin{equation}
\Delta = \sum_{j=1}^{m} v_j \cdot (k_j' - k_j),
\end{equation}
\begin{table*}[]
\centering
\caption{SAT Solver Performance Metrics for Various Benchmarks with \textit{LCTL} and \textit{CRTL} Logic Locking Configurations.}
\label{tab:sat}
\scalebox{0.95}{\begin{tabular}{ccccccccccccccc}
\hline
\multirow{2}{*}{\textbf{Circuit}} &
  \multirow{2}{*}{\#\textbf{Keys}} &
  \multirow{2}{*}{\textbf{Percent}} &
  \multirow{2}{*}{\textbf{Conflicts}} &
  \multirow{2}{*}{\textbf{Decisions}} &
  \multirow{2}{*}{\textbf{CPU Time}} &
  \multirow{2}{*}{\textbf{Result}} &
  \multicolumn{3}{c}{\textbf{\textit{LCTL}}} &
  \multicolumn{3}{c}{\textbf{\textit{CRTL}}} \\ \cline{8-13} 
      &    &    &     &       &          &         & A (\textmu $\text{m}^2$)   & P (\textmu $\text{W}$)  & D (ns)   & A (\textmu $\text{m}^2$)    & P (\textmu $\text{W}$)   & D (ns)   \\ \hline
\texttt{c1355} \cite{iscas85} & 16 & 50 & --- & ---   & ---      & \texttt{Timeout} & 380 & 150 & 60 & 280 & 120 & 25 \\
\texttt{c17} \cite{iscas85}  & 17 & 50 & 5   & 138   & 0.13 & SAT     & 5 & 2.5 & 1.0 & 3 & 1.2 & 0.5 \\
\texttt{c1908} \cite{iscas85} & 17 & 50 & --- & ---   & ---      & \texttt{Timeout} & 300 & 140 & 55 & 220 & 110 & 20 \\
\texttt{c2670} \cite{iscas85} & 17 & 50 & 0   & 0     & 16.7  & UNSAT   & 475 & 240 & 90 & 350 & 190 & 35 \\
\texttt{c7552} \cite{iscas85} & 17 & 50 & --- & ---   & ---      & \texttt{Timeout} & 1100 & 550 & 210 & 850 & 420 & 80 \\
\texttt{s1494} \cite{iscas89} & 16 & 20 & --- & ---   & ---      & \texttt{Timeout} & 350 & 170 & 60 & 260 & 130 & 22 \\
\texttt{s386} \cite{iscas89}  & 18 & 80 & 104 & 10797 & 37.9  & SAT     & 75 & 35 & 12 & 50 & 25 & 8 \\
\texttt{s526} \cite{iscas89}  & 18 & 80 & 109 & 19055 & 31.6  & SAT     & 95 & 47 & 18 & 70 & 35 & 12 \\
\texttt{s5378} \cite{iscas89} & 16 & 20 & 0   & 0     & 196.7  & UNSAT   & 820 & 400 & 160 & 600 & 300 & 60 \\
\texttt{s713} \cite{iscas89}  & 18 & 70 & 309 & 35636 & 34.1  & SAT     & 105 & 52 & 20 & 75 & 37 & 15 \\
\texttt{i10} \cite{yang1991logic}  & 15 & 50 & --- & ---   & ---      & \texttt{Timeout} & 540 & 360 & 400 & 450 & 300 & 150 \\
\texttt{i8} \cite{yang1991logic}  & 15 & 80 & --- & ---   & ---      & \texttt{Timeout} & 250 & 160 & 200 & 220 & 140 & 100 \\
\texttt{des*} \cite{yang1991logic}  & 10 & 5 & --- & ---   & ---      & \texttt{Timeout} & 9000 & 6000 & 700 & 7600 & 5000 & 300 \\
\texttt{b15*} \cite{davidson1999itc}  & 5 & 50 & --- & ---   & ---      & \texttt{Timeout} & 2300 & 1500 & 180 & 2000 & 1300 & 100 \\
\texttt{b17*} \cite{davidson1999itc}  & 10 & 10 & --- & ---   & ---      & \texttt{Timeout} & 7200 & 5000 & 550 & 6100 & 4000 & 250 \\\hline
\multicolumn{13}{l}{\small * The comparator of the \textit{LCTL} and \textit{CRTL} was modified to have 10\texttimes width of PMOS and NMOS for a faster and functional operation.} \\
\end{tabular}}
\end{table*}
which disrupts the threshold condition \( T \), obfuscating the circuit's behavior and enhancing security. These equations form the basis for integrating logic locking in threshold logic circuits. When applied to specific architectures like \textit{LCTL} and \textit{CRTL}, this approach ensures efficient and secure operation under varying design constraints

By leveraging the properties of the \textit{LCTL} and \textit{CRTL} architectures, we ensure seamless integration of key-driven security. Both architectures use a fixed-weight approach, where each input contributes uniformly to the sum. This uniformity simplifies the integration of key inputs while ensuring that any deviation from the correct key disrupts the threshold conditions. This not only obfuscates the circuit's functionality but also prevents unintended outputs, preserving the logical integrity of the design.

Consider a TLG with three input nodes \( x_1, x_2, x_3 \), weights \( w_1 = 1, w_2 = 1, w_3 = 1 \), and threshold \( T = 3 \):
\begin{equation}
\text{Output} = 1 \quad \text{if} \quad 1 \cdot x_1 + 1 \cdot x_2 + 1 \cdot x_3 \geq 3.
\end{equation}
With added key inputs \( k_1, k_2 \) and weights \( v_1 = -2, v_2 = 3 \), the equation becomes:
\begin{equation}
\text{Output} = 1 \quad \text{if} \quad 1 \cdot x_1 + 1 \cdot x_2 + 1 \cdot x_3 - 2 \cdot k_1 + 3 \cdot k_2 \geq 3.
\end{equation}
The correct key vector \( \mathbf{K} = [1, 1] \) neutralizes the effect of the key inputs, maintaining circuit functionality. Any incorrect key vector \( \mathbf{K}' \) results in a non-zero \( \Delta \), altering the circuit's output and ensuring security against unauthorized use.

\section{Results and Discussion}
\label{sec:results}
\subsection{SAT Attack Analysis}

The proposed logic locking mechanism was evaluated using \texttt{Minisat+} across benchmarks from ISCAS’85, ISCAS’89, ITC’99, and MCNC. Table~\ref{tab:sat} summarizes the SAT solver performance metrics for the two architectures under various locking configurations.

For complex circuits such as \texttt{c1355}, \texttt{c7552}, \texttt{i10}, and \texttt{b17}, the SAT solver consistently timed out under the one-hour timeout threshold set for all benchmarks. For instance, in the case of \texttt{i10}, the charge recycling-based design demonstrated superior efficiency by achieving lower area and delay compared to its latch-based counterpart, underscoring its scalability for larger designs.

Simpler circuits such as \texttt{c17} were solved efficiently, with the charge recycling approach outperforming across all performance metrics. This trend continued in more complex designs like \texttt{des} and \texttt{b15}, where the charge recycling configuration achieved significantly lower area and delay compared to conventional implementations.
\begin{figure}[htbp]
    \centering
    \includegraphics[width=0.4\textwidth]{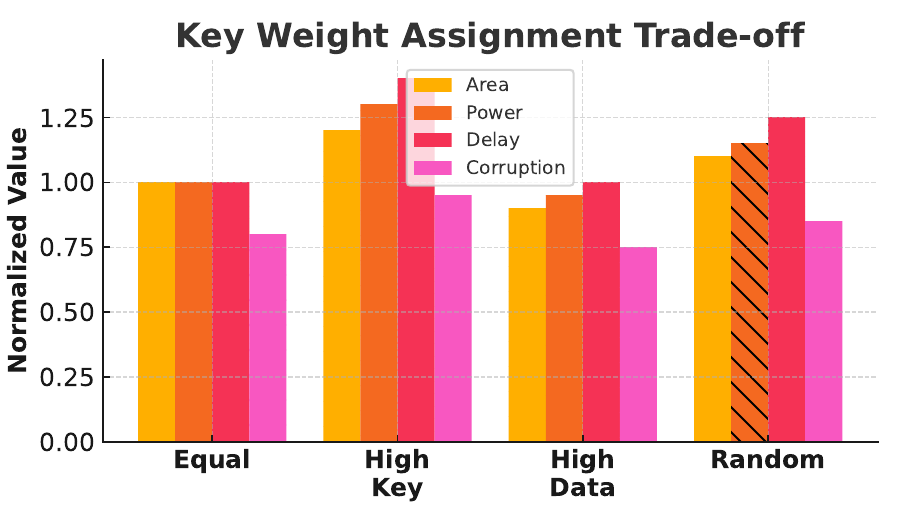}
    \caption{Impact of key input weight assignment on power, delay, and corruption rate in a TLG. Balanced weights offer higher corruption but incur power/delay trade-offs.}
   
    \label{fig:key_weight_tradeoff}
\end{figure}
Circuits with higher locking percentages, like \texttt{i8} at 80\%, exhibited increased SAT solver complexity. Similarly, in \texttt{b17}, with a moderate locking percentage, the charge recycling architecture consistently required less area and delay, reinforcing its suitability for high fan-in configurations. Modified circuits, such as \texttt{b15}, further emphasized this trend, where increased transistor widths for handling higher fan-in resulted in enhanced performance metrics in the recycling-based implementation.

\begin{figure}
    \centering
    \includegraphics[width=1\linewidth]{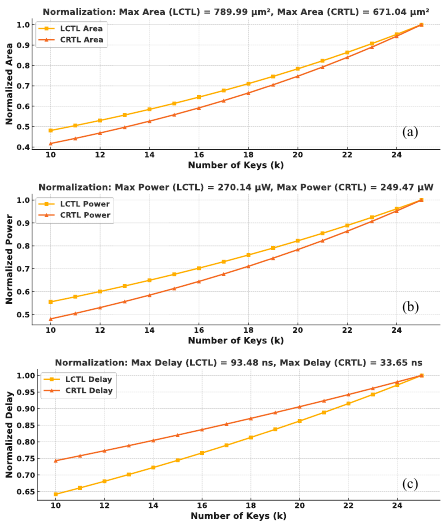}
    \caption{Normalized performance metrics a) Area, b) Power, and c) Delay overhead for latch-based (\textit{LCTL}) and recycling-based (\textit{CRTL}) architectures for \texttt{c1355} across varying key configurations (\(K = 10\) to \(K = 25\)).}
    \label{fig:comparison}
    \vspace{-2em}
\end{figure}

\subsection{Trade-Off Analysis of Key Weight Assignment}
\label{subsec:key_weight_tradeoff}

TLG-based logic locking introduces key inputs with weighted contributions toward the gate’s threshold, offering a richer design space than conventional binary keying. However, this flexibility introduces a crucial question: how should key weights be assigned to balance security (i.e., output corruption) against design overhead (i.e., power and delay)?

To explore this, we synthesized a representative 4-input TLG circuit and systematically varied the total cumulative weight of the key inputs. For each weight assignment, we measured dynamic power, propagation delay, and output corruption rate, defined as the percentage of incorrect primary outputs produced when incorrect keys are applied. A higher corruption rate indicates that wrong keys are more likely to cause functional errors in the circuit, making the design more resistant to SAT-based attacks. In this study, we evaluate the corruption rate by exhaustively simulating all possible incorrect key combinations for the TLG configuration and calculating the average output mismatch.
.

Figure~\ref{fig:key_weight_tradeoff} shows the results. Notably:
\begin{itemize}
    \item \textbf{Corruption rate} peaks at intermediate total key weights (e.g., 2–3), where key influence is strong enough to affect the output, yet not too dominant to bias it predictably.
    \item \textbf{Power and delay} increase linearly with key weight, reflecting greater switching activity and reduced noise margins in skewed TLG configurations.
\end{itemize}

To contextualize these trade-offs, Table~\ref{tab:normalized_comparison} provides a normalized comparison of our TLG-based locking scheme against conventional XOR and SFLL-based locking, across SAT attack time, area, power, and delay. Compared to XOR (which shows low overhead but weak SAT resilience) and SFLL (which increases overhead significantly to gain SAT resistance), our approach offers a tunable trade-off via key weight control.


\begin{table}[htbp]
\centering
\caption{Normalized Comparison of Locking Techniques}

\label{tab:normalized_comparison}
\begin{tabular}{lcccc}
\hline 
\textbf{Method} & \textbf{SAT Time} & \textbf{Area} & \textbf{Power} & \textbf{Delay} \\
\hline
XOR Locking      & 0.1× & 1.0× & 1.0× & 1.0× \\
SFLL-HD$^0$      & 1.0× & 2.2× & 1.9× & 1.8× \\
TLG (W=2–3)      & \textbf{3.0×} & \textbf{1.5×} & \textbf{1.2×} & \textbf{1.3×} \\
\hline
\end{tabular}
\vspace{-1em}
\end{table}

\subsection{Performance Analysis of TLG Architectures}

The comparison between the two TLG architectures, Latch-Controlled TLG (LCTL) and Charge-Recycling TLG (CRTL)—reveals distinct trade-offs, as illustrated in Fig.~\ref{fig:comparison}. The plots depict normalized area, power, and delay metrics across varying key sizes ($k$) for the \texttt{c1355} benchmark.

In terms of area, LCTL consistently incurs higher overhead. This is evident in benchmarks such as \texttt{s713} and \texttt{b17}, where LCTL exhibits significantly larger normalized values, due to its additional latch control complexity. In contrast, CRTL maintains a more compact area profile across all key sizes.

CRTL also demonstrates superior power efficiency. As indicated in benchmarks like \texttt{s386} and \texttt{i10}, it achieves lower normalized power consumption, attributed to the inherent energy-recovery mechanism in the charge-recycling network.

Delay analysis further favors the CRTL approach for high-speed applications. For high fan-in circuits such as \texttt{s526} and \texttt{b15}, CRTL achieves significantly reduced delay, making it ideal for timing-critical paths.

\section{Conclusion}
\label{sec:conclusion}

This work introduces a secure and efficient logic locking methodology that embeds key-driven obfuscation directly into Threshold Logic Gates (TLGs) synthesized from And-Inverter Graphs (AIGs). By leveraging the weighted sum computation inherent to TLGs, the proposed approach ensures correct circuit behavior only under the correct key, while resisting SAT-based and other logic unlocking attacks.

Comprehensive experiments on ISCAS’85, ISCAS’89, ITC’99, and MCNC benchmarks validate the effectiveness of this strategy. The Charge Recycling Threshold Logic (\textit{CRTL}) architecture consistently outperformed Latch-Type Low Power Threshold Logic (\textit{LCTL}) in delay and power efficiency—achieving up to 29\% lower power and 58\% faster delay. On large-scale designs like \texttt{b17}, \textit{CRTL} reduced area by 26\%, while retaining high security. On smaller benchmarks such as \texttt{s386} and \texttt{s526}, \textit{CRTL} achieved 25~μW power and 8~ns delay, demonstrating scalability.

Integrating logic locking into the TLG synthesis process reduces logic depth and improves attack resilience, while offering distinct trade-offs: \textit{CRTL} suits delay/power-constrained designs, and \textit{LCTL} is beneficial in area-sensitive contexts.


\bibliographystyle{IEEEtran}
\bibliography{references}

\end{document}